\global\def\draftcontrol{0}

   \def\versionno{  roll -- draft -- 3.04.03   }

\catcode`\@=11

\expandafter\ifx\csname draftcontrol\endcsname\relax\global\def\draftcontrol{0}
\fi

{\count255=\time\divide\count255 by 60
\xdef\hourmin{\number\count255}
\multiply\count255 by-60\advance\count255 by\time
\xdef\hourmin{\hourmin:\ifnum\count255<10 0\fi\the\count255}}
\def\draftdate{\number\month/\number\day/\number\year\ \ \ \hourmin }
 
\newcommand\makepapertitle{\par
  \begingroup
    \renewcommand\thefootnote{\@fnsymbol\c@footnote}%
    \def\@makefnmark{\rlap{\@textsuperscript{\normalfont\@thefnmark}}}%
    \long\def\@makefntext##1{\parindent 1em\noindent
            \hb@xt@1.8em{%
                \hss\@textsuperscript{\normalfont\@thefnmark}}##1}%
     \newpage
     \global\@topnum\z@   
     \@makepapertitle
     \thispagestyle{empty}\@thanks
  \endgroup
  \setcounter{footnote}{0}%
  \global\let\thanks\relax
  \global\let\makepapertitle\relax
  \global\let\@makepapertitle\relax
  \global\let\@thanks\@empty
  \global\let\@author\@empty
  \global\let\@date\@empty
  \global\let\@title\@empty
  \global\let\title\relax
  \global\let\author\relax
  \global\let\date\relax
  \global\let\and\relax
  \def\version{\let\version\@version\@gobble}
}
\def\@makepapertitle{%
  \newpage
   \ifnum\draftcontrol=1 {}
   \version\versionno
   \vskip 5em%
   \else
   \hfill\hbox to 3cm {\parbox{4cm}{\@pubnum}\hss}%
   \vskip 5em%
   \fi
   \begin{center}%
   \let \footnote \thanks
      {\hskip -0\textwidth \hbox to 1\textwidth%
        {\centerline{\Large\bf{\noindent\@title}}}}%
     \vskip 1.5em%
     {\normalsize
       \lineskip .5em%
       \begin{tabular}[t]{c}%
         \@author
       \end{tabular}\par}%
     \vskip 1em%
     {\@bstract}%
     \end{center}%
     \vskip 1em
     \@date%
   \par
}

\gdef\@pubnum{}
\def\pubnum#1{%
  \gdef\@pubnum{#1}}

\gdef\@bstract{}
\def\Abstract#1{%
  \gdef\@bstract{%
   \parbox{\textwidth-0pc}{%
   \centerline{\bf Abstract}\penalty1000
   \noindent
   \renewcommand\baselinestretch{1.0}
   {#1}}}
}

\def\ps@paper{\let\@mkboth\@gobbletwo%
     \ifnum\draftcontrol=1
        \def\@oddfoot{\hbox to \textwidth{\tiny \versionno \hfil\tiny\draftdate}%
        \hskip -\textwidth \hbox to \textwidth{\hfil\rm\thepage\hfil}}%
     \else\def\@oddfoot{\hbox to \textwidth{\hfil\rm\thepage\hfil}}
     \fi
     \let\@evenfoot\@oddfoot
}

\def\body{\clearpage
          \pagestyle{paper}
        }
\newenvironment{acknowledgments}{%
\vskip 3.25ex
\noindent {\bf Acknowledgments}
}

\def\@version#1{\ifnum\draftcontrol=1
\typeout{}\typeout{#1}\typeout{}
\vskip3mm\centerline{\hbox{\fbox{\normalsize{\tt DRAFT -- #1 -- }
                   {\draftdate}}}}\vskip3mm
\fi}
\let\version\@version
\long\def\eqlabel#1{\ifnum\draftcontrol=1
                    \tag@false  
                    \tag*{(\theequation) \hbox to -0.2cm{\hspace{0cm}\small{#1}\hss}}
                    \refstepcounter{equation} 
                    \edef\@currentlabel{\theequation}
                    \ltx@label{#1}          
                    \else
                    \label{#1}
                    \fi
                    }
\let\st@bibitem\@bibitem
\let\st@lbibitem\@lbibitem
\ifnum\draftcontrol=1
  \def\@bibitem#1{%
    \st@bibitem{#1}\a@@label{#1}\ignorespaces}
  \def\@lbibitem[#1]#2{%
    \st@lbibitem[#1]{#2}\a@@label{#2}\ignorespaces}
  \def\a@@label#1{%
    \gdef\a@lab{\smash{\normalfont\small#1}}
    \ifvmode
      \if@inlabel
        \global\setbox\@labels\hbox{%
          \llap{\a@lab\let\a@lab\relax
                \kern\@totalleftmargin\kern\marginparsep}%
          \box\@labels}%
      \fi
    \fi}
\fi

\documentclass[12pt,letterpaper]{article}

\usepackage{amsmath,amssymb,array,calc,amsthm,rotating,epsfig,psfrag}
\usepackage[nosort]{cite}

\renewcommand\baselinestretch{1.25}
\renewcommand\section{\@startsection {section}{1}{\z@}%
                                   {-3.5ex \@plus -1ex \@minus -.2ex}%
                                   {2.3ex \@plus.2ex}%
                                   {\normalfont\large\bfseries}}
\renewcommand\subsection{\@startsection{subsection}{2}{\z@}%
                                   {-3.25ex\@plus -1ex \@minus -.2ex}%
                                   {1.5ex \@plus .2ex}%
                                   {\normalfont\normalsize\bfseries}}
\renewcommand\subsubsection{\@startsection{subsubsection}{3}{\z@}%
                                   {-3.25ex\@plus -1ex \@minus -.2ex}%
                                   {1.5ex \@plus .2ex}%
                                   {\normalfont\normalsize\it}}
\renewcommand\paragraph{\@startsection{paragraph}{4}{\z@}%
                                   {-3.25ex\@plus -1ex \@minus -.2ex}%
                                   {1.5ex \@plus .2ex}%
                                   {\normalfont\normalsize\bf}}
\renewcommand\subparagraph{\@startsection{subparagraph}{5}{\z@}%
                                   {-1.25ex\@plus -1ex \@minus -.2ex}%
                                   {0ex \@plus .2ex}%
                                   {\normalfont\normalsize\it}}

\def\ie{{\it i.e.}}
\def\eg{{\it e.g.}}

\def\revise#1       {\raisebox{-0em}{\rule{3pt}{1em}}%
                     \marginpar{\raisebox{.5em}{\vrule width3pt\
                     \vrule width0pt height 0pt depth0.5em
                     \hbox to 0cm{\hspace{0cm}{%
                     \parbox[t]{4em}{\raggedright\footnotesize{#1}}}\hss}}}}

\def\calc         {{\cal C}}
\def\cald         {{\cal D}}

\def\calh         {{\cal H}}

\def\calo         {{\cal O}}

\def\complex      {{\mathbb C}}

\def\reals        {{\mathbb R}}
\def\zet          {{\mathbb Z}}

\def\del          {\partial}

\def\ee           {{\it e}}
\def\ii           {{\it i}}

\def\tr           {\mathop{\rm tr}}
\def\Re           {{\rm Re\hskip0.1em}}
\def\Im           {{\rm Im\hskip0.1em}}

\def\const        {{\it const.\,}}

\def\sqr#1#2{{\vcenter{\vbox{\hrule height.#2pt  
 \hbox{\vrule width.#2pt height#1pt \kern#1pt
 \vrule width.#2pt}\hrule height.#2pt}}}}

\newcommand{\bra}[1]     {{\langle#1|}}
\newcommand{\ket}[1]     {{|#1\rangle}}
\newtheorem{theorem}{Theorem}
\newtheorem{proposition}{Proposition}

\def\PI{$\hat{\text{\raisebox{.65em}{\rule{.95em}{.05em}}\hskip -1em PI}}$}

\catcode`\@=12

\begin{document}


\title{The Decay of Unstable Noncommutative Solitons$^{*}$}

\pubnum{%
NSF-ITP-03-05 \\
hep-th/0301119}
\date{January 2003}

\author{Thomas Chen$^{\%}$, J\"urg Fr\"ohlich$^{\&}$, 
and Johannes Walcher$^{@}$ \\[0.4cm]
\it ${}^{\%}$Courant Institute of Mathematical Sciences \\
\it New York University \\
\it New York, NY 10012-1185, U.S.A. \\[0.2cm]
\it ${}^{\&}$Institute for Theoretical Physics \\
\it ETH H\"onggerberg \\
\it 8093 Z\"urich, Switzerland \\[0.2cm]
\it ${}^{@}$Kavli Institute for Theoretical Physics \\
\it University of California \\
\it Santa Barbara, CA 93106, U.S.A.\\[0.4cm]
}

\Abstract{
We study the classical decay of unstable scalar solitons in
noncommutative field theory in $2+1$ dimensions. This can, but does 
not have to, be viewed as a toy model for the decay of D-branes in 
string theory. In the limit that the noncommutativity parameter $\theta$
is infinite, the gradient term is absent, there are no propagating 
modes and the soliton does not decay at all. If $\theta$ is large, but 
finite, the rotationally symmetric decay channel can be described as a 
highly excited nonlinear oscillator weakly coupled to a continuum of 
linear modes. This system is closely akin to those studied in the 
context of discrete breathers. We here diagonalize the linear problem 
and compute the decay rate to first order using a version of Fermi's 
Golden Rule, leaving a more rigorous treatment for future work. 
\vskip 1em
\noindent $^{*}$ Dedicated to Rudolf Haag, on the occasion of his
eightieth birthday, with admiration and affection
}

\enlargethispage{1em}

\makepapertitle


\body

\version\versionno

\section{Introduction}

Solitons are interesting objects. Originally discovered in experiments on 
shallow water waves one and a half centuries ago, they have since then 
become a rich and fascinating area of theoretical and mathematical study. 
Today, solitons are playing an increasingly important role in modern field 
and string theories.

A general issue in the analysis of solitons is stability, in both the 
dynamical (small perturbations of the initial conditions within a theory), 
and the structural (small deformations of the theory) sense. Stability can 
often be established on the basis of topological properties of the solution, 
or by the integrability of the underlying theory. On the other hand, a 
problem that has attracted much less attention concerns the dynamical long-time 
evolution of decaying, {\it unstable} solitons. The reason is apparently not 
because this question fails to be interesting, but rather because it is hard.

In this paper, we study the decay dynamics of a particularly simple class 
of solitons that were discovered only very recently, coined noncommutative 
solitons \cite{gms}. The corresponding model can be derived from the action 
functional
\begin{equation}
S = \int dx\, dy\, dt\; \Bigl[ {\frac12}(\del_t \Phi)^2 - 
{\frac12}(\nabla\Phi)^2 - V(\Phi) \Bigr] \,,
\eqlabel{action}
\end{equation}
which describes a real scalar field $\Phi(x,y,t)$ in $2+1$ dimensions.
Noncommutativity is introduced in the multiplication rule for $\Phi$,
as we review below. We will only consider space-space noncommutativity
in this paper, governed by a single noncommutativity parameter
$\theta\in[0,\infty)$.

``Noncommutative field theories'' of this kind appear for instance as 
world-volume theories of D-branes in string theory by turning on a large 
Neveu-Schwarz-Neveu-Schwarz B-field. In general, the resulting theory is 
more complicated than \eqref{action}, and will include gauge fields and so 
on, but here, we will restrict ourselves to \eqref{action}. We will not 
indulge here in a general discussion of noncommutative geometry \cite{connes}, 
and its role in modern physics. As starting points for references and 
citations we refer to the short list \cite{witten85,connesbook,matrix,cds,
sewi}, as well as to the reviews \cite{fgr,done,szabo}.

\subsection{Motivation from string theory}

Our interest in this problem arose from recent studies of tachyon 
condensation in string theory, hence let us briefly review this background
motivation here. Tachyon condensation in string theory is formalized 
in Sen's conjectures \cite{sen1,sen2}. Roughly, given a string theory 
background with an unstable D-brane or D-brane system, \ie, given a 
solution of string theory with an open string tachyon, one should look 
for another solution of string theory that is related to the first 
solution by ``tachyon condensation''. This solution corresponds to a 
different background with other branes, and can either be stable or 
contain other tachyons. In bosonic string theory, all D$p$-branes are 
unstable, and open string theory is conjectured to have a solution 
corresponding to the condensation of all these tachyons, characterized 
by:
\begin{itemize}
\item[(S1)] There are no perturbative open string excitations around
this solution, which is hence also known as the ``closed string vacuum of
open string field theory''.
\item[(S2)] The energy difference between the solution with open strings
and the closed string vacuum is exactly equal to the D-brane tension computed
in perturbative string theory.
\end{itemize}
(To avoid confusion, we note that the bosonic string in $26$ dimensions also 
has a tachyon in the closed string sector, whose role is, however, much less 
understood.)

A useful analogy in this context is to think of D-branes as 
``tachyonic solitons''. For example, a codimension-one brane in bosonic 
string theory is analogous to the solitonic lump solution in a scalar 
$\phi^3$ theory. This soliton and the D-brane are of course unstable. 
In superstring theory, a codimension-one brane is analogous to the stable 
solitonic kink in a scalar field theory with a $\phi^4$ double-well 
potential. The topological charge of such a soliton that guarantees 
stability is analogous to the Ramond-Ramond charge of a brane. These 
analogies break down at some level because string field theory contains 
many more fields than just the tachyon, and they are all excited in the 
vacua representing the endpoint of tachyon condensation. 

Tachyon condensation can be formulated on the worldsheet (as worldsheet 
RG flow) or in spacetime in open string field theory \cite{senzwi}. This 
has been quite successful and can be regarded as one of the major 
advances in the field over the last few years. Until recently, however, 
the research has focused on finding the endpoint of the condensation, and 
not on the decay of the unstable brane (the condensation of the tachyon) 
as a dynamical process in time.

Very recently, Sen \cite{sen10,sen11,sen12} has made precise proposals
for the construction of the time-dependent solutions corresponding to
the condensation process. This is based mainly on a worldsheet analysis,
and is motivated in part by the S-brane proposal of Gutperle and 
Strominger \cite{gust}. Again, there are many different possible processes
one can study. The generic features characterizing the solutions are
\begin{itemize}
\item[(S3)] If the initial configuration is an unstable brane configuration
of non-zero codimension (say a D$p$-brane in bosonic string theory with
$p<25$) then the energy stays localized in the plane of the original
brane, in other words, energy is not carried away from this plane by
classical radiation.
\item[(S4)] When computing the energy-momentum tensor of the solution
on finds that, asymptotically for late times, the energy density of 
the solution is constant (in the plane of the original brane) and the 
pressure vanishes exponentially in time.
\end{itemize}

It is important to emphasize that these are supposed to be {\it classical} 
solutions of string theory. In particular, the radiation in (S3) are 
open strings. It is very likely that in fact closed strings (which appear 
at one loop level in open string theory) do carry away energy 
\cite{gust,blw,strominger}. Furthermore, one may note that (S3) is in 
some sense a consequence of (S1). Namely, if one thinks of a 
lower-dimensional brane as a tachyonic soliton, then far away this 
solution looks like the closed string vacuum which does not allow 
open string excitations. Hence, there are no modes that could carry 
away energy. Also, (S2) and (S4) are of course partly trivial consequences 
of energy conservation.

All these questions are rather difficult to study in open string field 
theory. The underlying idea behind the present work was to study the 
dynamics of such decays in the so-called noncommutative (NC) field 
theory limit of string theory, in the hope that one could reproduce 
some of the above properties in a simple manner. 

The reason for this hope is the fact that solitons in noncommutative 
field theories (NC solitons) have many properties in common with D-branes 
as tachyonic solitons. NC solitons and instantons in gauge theories 
have been studied independently of strings for quite some time, most 
notably by Nekrasov and Schwarz \cite{nesc}, and more recently by 
Harvey et al. \cite{hklm}, Gross and Nekrasov \cite{grne,grne2}, 
Polychronakos \cite{poly}, and others. In NC scalar field theories, 
solitons were discovered by Gopakumar, Minwalla and Strominger \cite{gms}. 
For a summary of the relation to D-branes, in particular as far as their 
charges and descent relations are concerned, see Harvey's lectures 
\cite{harvey}.

An explanation of why NC solitons (in the infinite $\theta$ limit)
are useful for describing tachyon condensation in string theory was 
given by Witten \cite{witten73}. The point of \cite{witten73} is that in 
the Seiberg-Witten limit, the open string field algebra factorizes into 
the Moyal algebra built on tachyon vertex operators and another algebra 
that ``contains all the stringy mystifications''. Therefore, in describing 
tachyon condensation, one can concentrate on the tachyon, the remaining 
pieces being added ``trivially'' at the end. 

Given all this, it becomes a natural problem, from the point of view
of string theory, to study the decay of unstable NC solitons, using
(S1)-(S4) as guiding questions, as a toy model for the dynamics of
open string tachyons on unstable D-branes.

\subsection{Summary of Results}

We will see that it is in fact not possible to reproduce the 
qualitative behavior described above in the simple model 
\eqref{action}. A priori, one would not expect (S3) simply
because of the failure of (S1) in the context of NC tachyon
condensation. Indeed, while NC solitons have localized energy,
far away the field is in the usual vacuum that has local
propagating excitations. However, we note that intuitively
(S1) is sufficient but not necessary for (S3), and in this
sense, finding (S3) would have been a nice substitute for (S1).
And recall that the very existence of NC solitons at finite
$\theta$ relies on the fact that the kinetic term is a small
perturbation that does not destroy the qualitative properties
at infinite $\theta$. One could expect something similar in the 
time-dependent case.

What we will find is that, except for infinite $\theta$, the
energy that initially is stored in the soliton does not stay
localized, but is carried away to infinity in the form of
classical radiation. Nevertheless, the qualitative properties of 
the decay do depend on the noncommutativity parameter $\theta$. In 
particular, the decay rate is exponentially suppressed in $\theta$, 
for large $\theta$.

We will proceed as follows. We start with a review of NC scalar
solitons and their basic properties in section \ref{ncft}.
It turns out that restricted to rotationally symmetric configurations,
the system is similar to certain nonlinear lattices which exhibit 
the phenomenon of ``discrete breathers'', and we review those in 
section \ref{discrete} for completeness. We will then present 
some numerical results in section \ref{numerics} in order to 
illustrate the qualitative properties of the decay. In section 
\ref{estimate1}, we turn to reproducing this behavior through an 
approximate calculation of the decay rate. The technical core is 
here the explicit diagonalization of the linear problem in 
section \ref{A}, that puts us in a position to give a closed 
expression for the estimated decay rate in section \ref{estimate2}. 
The approximation method relies on a certain version of Fermi's 
Golden Rule that we will explain.

To conclude this introduction, we note that a different toy model
for tachyon condensation in string theory, based on conventional
field theory, has been considered by Zwiebach \cite{zwiebach30}. It 
would be interesting to compare the dynamics of the decay with our
present results. Also, Moeller and Zwiebach \cite{mozw} have studied 
the time evolution of decaying solitons in $p$-adic string theory. 
The relation to our work is not clear. Finally, we note that some of 
the problems that we study here have also been broached in \cite{acso} 
and \cite{acatrinei}. The stability of non-commutative scalar 
solitons is also discussed in \cite{jackson,rocek}.

\section{Noncommutative scalar field theory in $2+1$ dimensions and
its solitons}
\label{ncft}

The theory of our interest \eqref{action} is defined over a model of 
$2+1$-dimensional spacetime in which time is an ordinary variable and 
space is the two-dimensional quantum mechanical phase space of a 
nonrelativistic particle in one dimension. In other words, we base space
on the Heisenberg-Born-Jordan commutation relation, 
\begin{equation} 
[x,y] = \ii\theta \,. 
\eqlabel{nc} 
\end{equation} 

It is well-known since the times of Weyl, Wigner, and Moyal that the algebra 
of functions on this space is a deformation of the ordinary algebra 
$\calc(\reals^2)$ along the parameter $\theta$. In one particular ordering
prescription, the product can be written as
\begin{equation}
\Phi_1\star\Phi_2 (x,y) = \ee^{\ii \frac\theta2(\del_{\xi_1}\del_{\eta_2} - 
\del_{\xi_2}\del_{\eta_1})} 
\Phi_1(\xi_1,\eta_1) \Phi_2(\xi_2,\eta_2) |_{(\xi_1,\eta_1)=(\xi_2,\eta_2)=(x,y)}\,.
\eqlabel{moyal}
\end{equation} 

Upon representing the commutation relations \eqref{nc} in the usual manner on
Hilbert space $\calh$, eq.\ \eqref{moyal} turns into a prescription for 
translating a function over phase space, such as our field $\Phi$, into a linear 
operator on $\calh$. Moreover, derivatives become commutators,
\begin{equation}
\del_x \longleftrightarrow \frac\ii\theta [y,\cdot]\,, \qquad\qquad
\del_y \longleftrightarrow -\frac\ii\theta [x,\cdot] \,,
\eqlabel{derivatives}
\end{equation}
and the integral a trace,
\begin{equation}
\int dx\, dy \longleftrightarrow 2\pi\theta\tr \,.
\eqlabel{trace}
\end{equation}
This yields, after setting
\begin{equation}
a = \frac 1{\sqrt{2\theta}} (x+\ii y)\,, \qquad\qquad
a^\dagger = \frac 1{\sqrt{2\theta}} (x-\ii y) \,,
\eqlabel{osci}
\end{equation}
the following expression for $S$.
\begin{equation}
S = 2\pi\theta \int dt \tr\Bigl[ {\frac12} (\del_t\Phi)^2
- {\frac{1}{\theta}} [a,\Phi][\Phi,a^\dagger] - V(\Phi) \Bigr]
\eqlabel{actop}
\end{equation}
We also record the equations of motion
\begin{equation}
\del_t^2 \Phi +\frac 2\theta [a,[a^\dagger,\Phi]] + V'(\Phi) = 0 \,.
\eqlabel{eom}
\end{equation}

\subsection{Rotational symmetry}

These expressions simplify further on rotationally symmetric field 
configurations,
\begin{equation}
\Phi(t) = \sum_{n=0}^\infty \lambda_n(t) P_n \,.
\eqlabel{ansatz}
\end{equation}
Here, $P_n=\ket n\bra n$ is the projector onto the $n$-th excited state
in the harmonic oscillator basis for $\calh$. One finds
\begin{equation}
[a,[a^\dagger,P_n]] = (2n+1) P_n - (n+1) P_{n+1} - n P_{n-1}\,.
\end{equation}
The equations of motion \eqref{eom} become
\begin{equation}
\begin{split}
\ddot\lambda_n & = \frac 2\theta \bigl((n+1)(\lambda_{n+1}-\lambda_n)
- n(\lambda_n - \lambda_{n-1}) \bigr) - V'(\lambda_n) \\
\ddot\lambda_0 & = \frac 2\theta (\lambda_1-\lambda_0) - V'(\lambda_0) \,.
\end{split}
\eqlabel{eqmot}
\end{equation}
This is a hamiltonian system with Hamiltonian
\begin{equation}
H = 2\pi\theta\sum_{n=0}^{\infty} \bigl[ {\frac12} \kappa_n^2
+ \frac n\theta (\lambda_n-\lambda_{n-1})^2 + V(\lambda_n)
\bigr] \,,
\eqlabel{hamilton}
\end{equation}
where $\kappa_n$ are the momenta canonically conjugate to the $\lambda_n$.

To connect \eqref{eqmot} and \eqref{hamilton} to something more familiar,
we note that using \eqref{moyal} to write the projector $P_n$ as a function 
on $\reals^2$, it has a peak around
\begin{equation}
r^2 = \bra n x^2 + y^2 \ket n \approx 2\theta n \,,
\eqlabel{rton}
\end{equation}
and that substituting \eqref{rton} into the radial part of the Laplace 
operator written in two-dimensional polar coordinates, one obtains
\begin{equation}
\frac 1r \del_r r\del_r = \frac 2\theta \del_n n \del_n \,.
\eqlabel{rlapl}
\end{equation}
Upon discretizing $n$ and writing $\del_n$ as a difference operator, one
reproduces \eqref{eqmot} or \eqref{hamilton}.

In this paper, we study properties of the dynamical system \eqref{eqmot}.

\subsection{Non-commutative solitons}

We start with static solutions of \eqref{eom} and \eqref{eqmot}, the 
noncommutative solitons. It was pointed out in \cite{gms} that in 
the limit $\theta\to\infty$, eq.\ \eqref{eom} simply reduces to 
$V'(\Phi)=0$ and can be solved by a projector $\Phi=\lambda P$ with
$\lambda\in\reals$ such that $V'(\lambda)=0$. One then treats 
the kinetic term as a perturbation and shows that the solutions 
continue to exist for $\theta<\infty$. Since, according to Derrick's 
theorem, the solutions do not exist in the classical or ``continuum'' 
limit $\theta\to 0$, there must be a critical $\theta_c$ beyond which 
the NC solitons cease to exist. Of course, existence is not controlled 
by $\theta$ alone, but rather by certain dimensionless quantities built 
out of $\theta$ and the parameters of $V$. 

The NC scalar solitons which are related to projectors in the 
$\theta\to\infty$ limit are referred to as the GMS 
(Gopakumar-Minwalla-Strominger) solitons. Instead of reviewing the 
large number of works devoted to GMS solitons, let us quote an 
existence theorem proved in \cite{djn2}, see also \cite{djn}. 

\begin{theorem}
\label{existence}
For every sequence $\lambda^{(0)}=\{\lambda_0^{(0)},\lambda_1^{(0)},\ldots\}$ 
with $V'(\lambda_n^{(0)})=0$ and $V''(\lambda_n^{(0)})>0$ for all $n$ (\ie, 
a stable static solution of \eqref{eqmot} at $1/\theta=0$), there exists a 
unique continuation $\lambda(\theta)=\{\lambda_0(\theta), \lambda_1(\theta),
\ldots\}$ to static solutions of \eqref{eqmot} for $0\le 1/\theta < 
1/\theta_c$, with some $\theta_c$ which depends on the chosen initial 
sequence $\lambda^{(0)}$.
\end{theorem}

\begin{proof}[We sketch the proof given in \cite{djn2}.]
We work on the Hilbert space $l_2(\zet_{\ge0})=\calh=
\{\lambda;\sum_{n=0}^\infty |\lambda_n|^2<\infty\}$. Let 
$(A\lambda)_n =-(n+1)(\lambda_{n+1}+\lambda_n)-n(\lambda_n-
\lambda_{n-1})$ be the discretized radial part of the Laplacian 
according to \eqref{rlapl}, with domain $\cald$.

The equation we have to solve is then
\begin{equation}
\epsilon A \lambda + V'(\lambda) = 0 \,,
\eqlabel{fix}
\end{equation}
where $\epsilon=2/\theta$. 

The first step of the proof is to show that the operator
$\epsilon A + V''(\lambda^{(0)})$ is invertible with
bounded inverse as a map from $\cald$ to $\calh$. This essentially
follows from $V''(\lambda^{(0)})>0$.

One then rewrites \eqref{fix} as
\begin{equation}
\lambda = \bigl(\epsilon A +V''(\lambda^{(0)})\bigr)^{-1}
\bigl[V''(\lambda^{(0)})\lambda^{(0)} + V'(\lambda^{(0)})-
V'(\lambda) - V''(\lambda^{(0)})(\lambda^{(0)}-\lambda)
\bigr] =:T_\epsilon(\lambda) \,,
\end{equation}
and shows that $T_\epsilon:\calh\to\calh$ is a contraction in a
neighborhood of $\lambda^{(0)}$ (for reasonable regularity assumptions
on $V$). The Banach fixed point theorem then implies the existence of 
a solution, with smooth dependence on $\epsilon$ in the appropriate
norms.
\end{proof}

\subsection{Stable and unstable solitons}

In order to simplify the following discussion, we assume that the 
potential $V$ is such that it has one global minimum $\lambda_{\it MIN}$, 
one local minimum $\lambda_{\it min}\neq \lambda_{\it MIN}$, and one 
local maximum $\lambda_{\it max}$. Corresponding statements should
hold for a general potential. 

In ref.\ \cite{djn2}, certain statements are proved concerning 
the stability of the solutions whose existence for $1/\theta>0$ is 
established by the above Theorem \ref{existence}. Stability is here 
defined as positivity of the Hessian of the Hamiltonian 
$H(\kappa_n=0,\lambda)$, eq.\ \eqref{hamilton}, viewed as a quadratic 
form. For example, if $V(\lambda)$ is as decribed above, then 
the solutions obtained as continuations of $\lambda^{(0)}$ to 
$1/\theta>0$ are stable if and only if $\lambda^{(0)}$ is of the form 
$\lambda_0^{(0)}=\lambda_1^{(0)}=\cdots=\lambda_N^{(0)}= 
\lambda_{\it min}$ for some $N$ and $\lambda_n^{(0)}=\lambda_{\it MIN}$ 
for $n>N$. 

It is interesting to note that the unstable direction of the unstable 
solitons of Theorem \ref{existence} (\eg, the one connected to 
$\lambda_0^{(0)}=\lambda_2^{(0)} =\cdots=\lambda_{\it MIN}$, $\lambda_1^{(0)} 
=\lambda_{\it min}$) is not rotationally symmetric. More precisely,
there exists a non-rotationally symmetric instability whenever the
sequence $\lambda_0(\theta),\lambda_1(\theta),\ldots\,$ at finite
$\theta$ is not monotonic. The nature of the instability implies that 
the decay of these solitons can not be studied with an ansatz of the form 
\eqref{ansatz}. 

Here, we shall be interested in a different kind of unstable soliton,
namely the one obtained from continuation to $1/\theta>0$ of the static 
unstable solution $\lambda_0^{(0)}=\lambda_{\it max}$, $\lambda_n^{(0)}=
\lambda_{\it MIN}$ for $n>0$. It is in fact rather simple to adapt the
above proof to show existence of these solitons as static solutions.%
\footnote{We thank M.\ Salmhofer for this suggestion.} The essential 
point is that $\epsilon A + V''(\lambda^{(0)})$ is still invertible even 
if $V''(\lambda_n^{(0)})$ is negative for a finite number of $n$'s, but
positive elsewhere. This is because it differs from a positive operator 
only by a finite rank perturbation, so that $\epsilon A+V''(\lambda^{(0)})$ 
can still be inverted for $\epsilon$ in a finite neighborhood of $0$.
This soliton is unstable, but the instability is rotationally symmetric.
Moreover, the sequence of $\lambda_n$'s is monotonic, so that at least
initially, the soliton does not suffer from the non-rotationally 
symmetric instability discussed above. This justifies an ansatz of the 
form \eqref{ansatz}.

Summarizing, we can imagine studying the decay and long time evolution 
of the following unstable initial conditions, slightly perturbing them away 
from the stationary point.
\begin{itemize}
\item[(I1)] A GMS soliton at finite $\theta$ obtained from an unstable
static solution at infinite $\theta$, say with $\lambda_0^{(0)} = 
\lambda_{\it max}$, $\lambda_n^{(0)}=\lambda_{\it MIN}$ for $n>0$.
\item[(I2)] A GMS soliton at finite $\theta$ obtained from a stable
configuration at infinite $\theta$, say $\lambda_0^{(0)}=
\lambda_2^{(0)}=\cdots=\lambda_{\it MIN}$, $\lambda_1^{(0)}=
\lambda_{\it min}$.
\item[(I3)] The unstable maximum, $\lambda_n=\lambda_{\it max}$
for all $n$.
\end{itemize}

The questions to ask about the decay are quite different from case to
case. In view of the relation between the noncommutative tachyon and
string field theory explained in \cite{witten73}, one can draw an
analogy between (I1) and the decay of an unstable D$23$ brane in
bosonic string theory, in which one looks at the directions 
{\it transverse} to the brane. In view of (S3), then, one can ask
if and how fast the energy that initially is localized is transported
away to infinity. We will see that while energy is indeed dissipating
whenever $1/\theta>0$, the decay is very slow.

We emphasize, again, that the action \eqref{action} is not a
good model for the tachyon of string field theory, even in the
large B-field limit. Rather, as explained in \cite{witten73}, not 
only does the algebra factorize, but also the equations of motion
do, and the tachyon has no dynamics in this picture. Thus, in the
large B-field limit, one always obtains the noncommutative ``field''
theory at infinite $\theta$. It is a nice, but rather trivial
observation of consistency that in this limit the energy does stay
localized \cite{acso}.

We find it harder to give an interpretation of the decays that start
from (I2). We will not study these decays here, but we expect that
some generalization of the technology that we develop could be
applied in that case.

We finally comment on the initial conditions (I3). Again by analogy
with string field theory, one might regard (I3) simply as the decay
of the open string vacuum (or of a lower dimensional D-brane in which
one looks in the directions {\it along} the brane). Then, one could 
ask for comparison with property (S4) of open string tachyons, in 
other words, can one see the formation of ``tachyon matter'' 
\cite{sen10,sen11,sen12}. We will not study this here, by let us 
observe that if the perturbation away from the stationary point is 
uniform, \ie, independent of $n$, then the equations of motion can 
be trivially solved to yield uniform oscillations. So, starting from 
(I3), one may want to ask what happens under {\it non-uniform 
perturbations}. It also makes little sense to restrict to rotationally 
symmetric decays in this case.

In this paper, we will only deal with initial conditions of the form (I1).

\section{Discrete breathers}
\label{discrete}

In order to explain some intuition about the behavior of the system
\eqref{eqmot}, we here digress on the theory of discrete breathers.

Discrete breathers are localized periodic solutions of lattices of 
oscillators, in which localization is not due to disorder (as in the
usual case of Anderson localization), but to nonlinearities. A useful
review is \cite{flwi}. The original discovery can be traced back to
\cite{tks,cape}, in which numerical simulations on these lattices
were performed. A rigorous existence proof was given in \cite{maau}, 
further and related earlier results appear in 
\cite{sema,mase,flwi,alfr,alfr2}.

As an example, let us consider a Hamiltonian similar to \eqref{hamilton},
\begin{equation}
H = \sum_{n=-\infty}^{\infty} \bigl[ \frac 12\kappa_n^2 + V(\lambda_n) + 
\epsilon_n (\lambda_n-\lambda_{n-1})^2 \bigr]\,.
\eqlabel{dbham}
\end{equation}
For $\epsilon_n=0$, this is just an infinite collection of independent
classical oscillators. For oscillations around a minimum 
$\lambda_{\it MIN}$ of $V$ (we have in mind a potential similar to the
one described in the previous section), we introduce canonical 
action-angle variables $(J_n,\phi_n)$, with
\begin{equation}
J_n = \frac1{2\pi}\oint \sqrt{2(E(J_n)-V(\lambda_n))}\;d\lambda_n \,,
\end{equation}
and we let $\phi_n=0$ correspond to the turning point of the oscillation, 
$\kappa_n=0$. These coordinates are valid until $\omega(J_n)=
\frac{dE(J_n)}{dJ_n}=0$. Such oscillations are localized at whichever 
site of the lattice is excited. 

On the other hand, if we couple the oscillators with $\epsilon_n=
\epsilon\neq 0$ but small and constant, the lattice has propagating 
small oscillations (the phonons). The spectrum of the linearized 
problem is 
\begin{equation}
\omega_{\it Ph}^2(k)=\omega_0^2 + 4\epsilon (1-\cos k) \,,
\end{equation}
where $\omega_0^2=V''(\lambda_{\it MIN})$, and $k\in[0,2\pi]$.

The linear spectrum is therefore bounded and given a nonlinear
oscillation of the problem at $\epsilon=0$ that is not in resonance 
with $\omega_0$, \ie, $m\omega(J_n)\neq \omega_0$ for all $m\in
\zet$, one can avoid resonances with the phonons, $m\omega(J_n)
\neq\omega_{\it Ph}(k)$ for all $m\in\zet$ and $k\in[0,2\pi]$, if
$\epsilon$ is sufficiently small. It is therefore plausible that
such an oscillation can continue to exist and be localized for 
$\epsilon\neq 0$. This intuitive argument can be turned into a 
rigorous existence theorem for discrete breathers \cite{maau}.
\begin{theorem}
\label{breather}
Every periodic solution of the decoupled nonlinear problem with
frequency $\omega$ satisfying the nonresonance condition 
$m\omega\neq\omega_0$ for all $m\in\zet$ and the anharmonicity 
condition $\frac{d\omega(J_n)}{dJ_n}\neq 0$ for all $n\in\zet$
and with initial conditions coded by a sequence $\eta_n\in\{0,1\}$, 
\begin{align}
\phi_n(0) &= 0    && \text{if $\eta_n=1$} \notag\\
\phi_n(0) &= \pi  && \text{if $\eta_n=-1$} \eqlabel{initial}\\
\lambda_n(0) &= \lambda_{\it MIN} && \text{if $\eta_n=0$} \notag \,,
\end{align}
can be uniquely continued to a periodic solution of the 
full nonlinear problem $\epsilon_n=\epsilon$, for $\epsilon\ge 0$
sufficiently small. These oscillations are localized in
the sense that $|\lambda_n(t)-\lambda_{\it MIN}|$ decays exponentially 
in $n$ whenever $\eta_n\to 0$.
\end{theorem}

The proof, which can be found in \cite{maau}, uses an implicit
function theorem and is rather elementary. A more geometric
formulation is given in \cite{sema}. The periodic solutions given
by the theorem are dynamically stable, in the sense that small
perturbations of the initial conditions remain small for all times.
This is discussed in an abstract setting in \cite{mase}.

We note that the solution that one wishes to continue has to be
strictly periodic at $\epsilon=0$, \ie, all excited oscillators 
have to be in resonance and they also have to be in phase, as
coded by \eqref{initial}. This resonance condition appears to be
opposite to the one found in the formulation of the KAM theorem.
However, it is completely in agreement with the intuitive argument.
If two frequencies are not rationally related, there will always
be a resonance of some higher harmonics with the phonon band,
leading to decay.

\paragraph{What can discrete breathers teach us about NC 
soliton decay?} 

The analogy  with discrete breathers leads to the question whether
breather-like solutions could exist in NC field theories. The answer
seems to be no.

The obvious difference is that in \eqref{hamilton} the coupling 
$\epsilon_n=n/\theta$ is not bounded as soon as $\theta<\infty$.
The linear spectrum, which will be discussed in detail in the next
section, is therefore unbounded. This leads to a failure of the
intuitive argument given above for existence of breathers. Any 
nonlinear oscillation will be in resonance with the phonons, no
matter how small $1/\theta$. There could be some hope if the
linear spectrum was pure point, as in Anderson localization. 
However, as we will see below, the spectrum is purely absolutely 
continuous.

It is a little more difficult to understand what goes wrong with the 
proof of \cite{maau}. But recall that one of the assumptions for the 
implicit function theorem is continuity of the derivative, and this 
seems to fail here, because the derivative with respect to $1/\theta$ 
is unbounded. While there might still be some hope that one could use 
slightly stronger versions of the implicit function theorem to 
prove existence, this will generically be restricted to very special 
conditions, and one cannot expect these breathers to be stable. 
In any case, our numerical results and the approximation computation 
of the decay rate clearly disfavor the existence of ``noncommutative 
breathers''.

\section{Numerical Results}
\label{numerics}

We now turn to presenting some results obtained by studying the
system \eqref{eqmot} using numerics on the computer. We use the 
potential
\begin{equation}
V(\Phi) = 4\Phi^2 (\Phi-2)^2
\eqlabel{pot}
\end{equation}
and initial conditions of the form (I1). Specifically, we take
\begin{equation}
\begin{split}
\lambda_0(0)&=\lambda_{\it max}=1 \\ 
\lambda_n(0)&=\lambda_{\it MIN}=0 \quad\; (n>0)
\end{split}
\qquad\qquad\dot\lambda_n(0)=0 \quad\;\text{(for all $n$)}\,.
\eqlabel{init}
\end{equation} 
For nonzero $1/\theta$, this is not a stationary point, and will
lead to some decay.

We have to work with a finite system $0\le n\le N$. Therefore, in order 
to see a decay, if it occurs, we have to allow for a leak of energy. We
do this by imposing ``absorbing boundary conditions'', \ie, the discrete 
version of
\begin{equation}
\frac{\del}{\del t} + \frac{1}{\sqrt{r}}\frac{\del}{\del r}\sqrt{r}=0\,,
\eqlabel{absorbing}
\end{equation}
(recall that outgoing spherical waves in two dimensions behave 
asymptotically like $\ee^{\ii E(t-r)}/\sqrt{r}$), which in view of 
\eqref{rton} reads
\begin{equation}
\dot\lambda_{N+1} + \sqrt{\frac{2(N+1)}\theta} (\lambda_{N+1}-\lambda_N)
+ \frac{1}{\sqrt{8\theta (N+1)}} \lambda_{N+1} =0  \,.
\eqlabel{bound}
\end{equation}

\begin{figure}[p]
\begin{center}
\psfrag{0.2}{$0.2$}
\psfrag{0.4}{$0.4$}
\psfrag{0.6}{$0.6$}
\psfrag{0.8}{$0.8$}
\psfrag{1}{$1$}
\psfrag{10}{$10$}
\psfrag{20}{$20$}
\psfrag{30}{$30$}
\psfrag{40}{$40$}
\psfrag{50}{$50$}
\psfrag{t}{$t$}
\psfrag{En}{$E_{\it loc}(t)/E_{\it loc}(0)$}
\epsfig{file=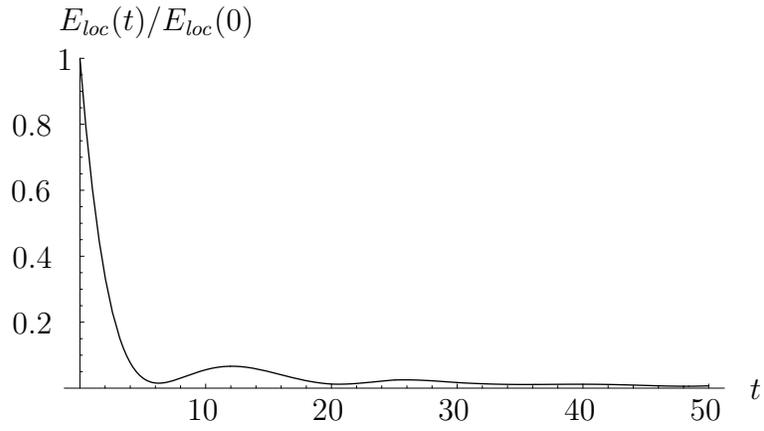,width=0.75\textwidth}
\caption{Energy \eqref{Eloc} stored in the first four oscillators of 
the system \eqref{eqmot}, truncated to $21$ oscillators with boundary 
conditions \eqref{bound} ($N=20$). The initial conditions are \eqref{init} 
for $n\le N$ and $\lambda_{N+1}(0)=0$, and $\theta=0.1$.}
\label{decay1}
\end{center}
\end{figure}

\begin{figure}[p]
\begin{center}
\psfrag{0.2}{$0.2$}
\psfrag{0.4}{$0.4$}
\psfrag{0.6}{$0.6$}
\psfrag{0.8}{$0.8$}
\psfrag{1}{$1$}
\psfrag{10}{$10$}
\psfrag{20}{$20$}
\psfrag{30}{$30$}
\psfrag{40}{$40$}
\psfrag{50}{$50$}
\psfrag{t}{$t$}
\psfrag{En}{$E_{\it loc}(t)/E_{\it loc}(0)$}
\epsfig{file=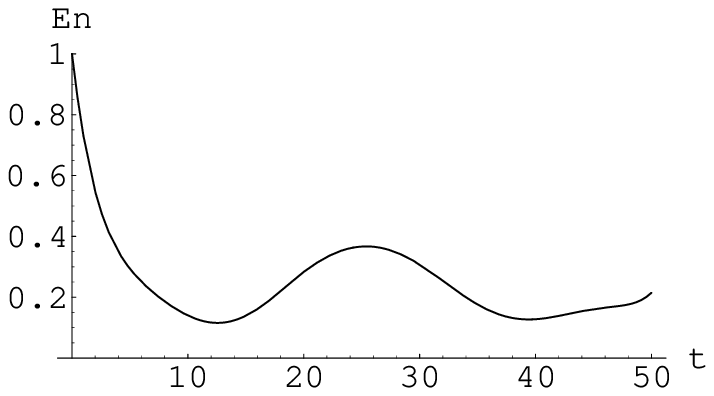,width=0.75\textwidth}
\caption{Same as fig.\ \ref{decay1}, for $\theta=0.2$.}
\label{decay2}
\end{center}
\end{figure}

\begin{figure}[p]
\begin{center}
\psfrag{0.2}{$0.2$}
\psfrag{0.4}{$0.4$}
\psfrag{0.6}{$0.6$}
\psfrag{0.8}{$0.8$}
\psfrag{1}{$1$}
\psfrag{10}{$10$}
\psfrag{20}{$20$}
\psfrag{30}{$30$}
\psfrag{40}{$40$}
\psfrag{50}{$50$}
\psfrag{t}{$t$}
\psfrag{En}{$E_{\it loc}(t)/E_{\it loc}(0)$}
\epsfig{file=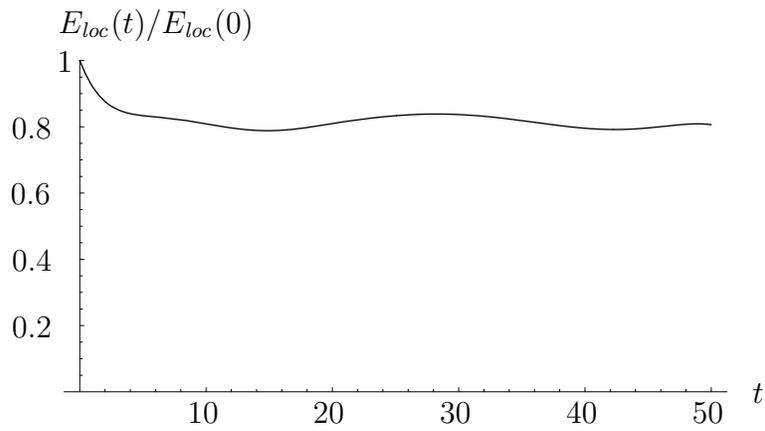,width=0.75\textwidth}
\caption{Same as fig.\ \ref{decay1}, for $\theta=0.35$.}
\label{decay3}
\end{center}
\end{figure}

\begin{figure}[p]
\begin{center}
\psfrag{0.2}{$0.2$}
\psfrag{0.4}{$0.4$}
\psfrag{0.6}{$0.6$}
\psfrag{0.8}{$0.8$}
\psfrag{1}{$1$}
\psfrag{10}{$10$}
\psfrag{20}{$20$}
\psfrag{30}{$30$}
\psfrag{40}{$40$}
\psfrag{50}{$50$}
\psfrag{t}{$t$}
\psfrag{En}{$E_{\it loc}(t)/E_{\it loc}(0)$}
\epsfig{file=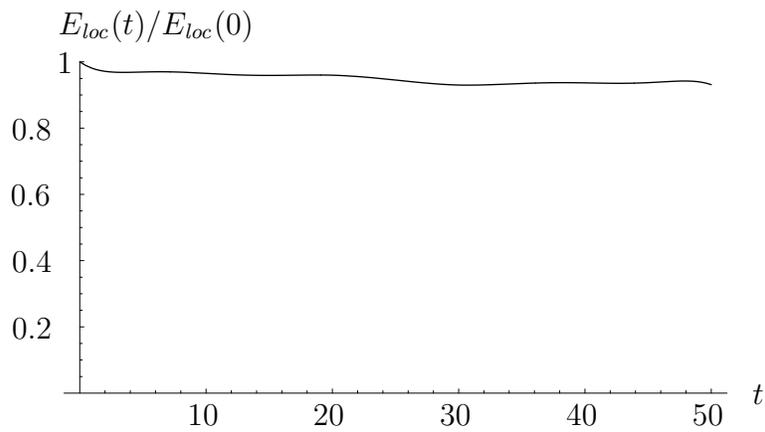,width=0.75\textwidth}
\caption{Same as fig.\ \ref{decay1}, for $\theta=0.7$.}
\label{decay4}
\end{center}
\end{figure}

We start by plotting, as a function of time, the energy $E_{\it loc}$
that is stored in the first four (say) oscillators, \ie,
\begin{equation}
E_{\it loc}(t) = \sum_{n=0}^3 \bigl[\frac 12\dot\lambda_n^2 + 
V(\lambda_n)+ \frac n\theta(\lambda_n-\lambda_{n-1})^2\bigr] \,,
\eqlabel{Eloc}
\end{equation}
divided by the initial energy, $E_{\it loc}(0)$. For different values of
$\theta$, our results are shown in figs.\ \ref{decay1}-\ref{decay4}.

\clearpage

\begin{figure}[h]
\begin{center}
\psfrag{0.2}{$0.2$}
\psfrag{0.4}{$0.4$}
\psfrag{0.6}{$0.6$}
\psfrag{0.8}{$0.8$}
\psfrag{1}{$1$}
\psfrag{theta}{$\theta$}
\psfrag{rat}{${\overline{E_{\it loc}}}/E_0$}
\epsfig{file=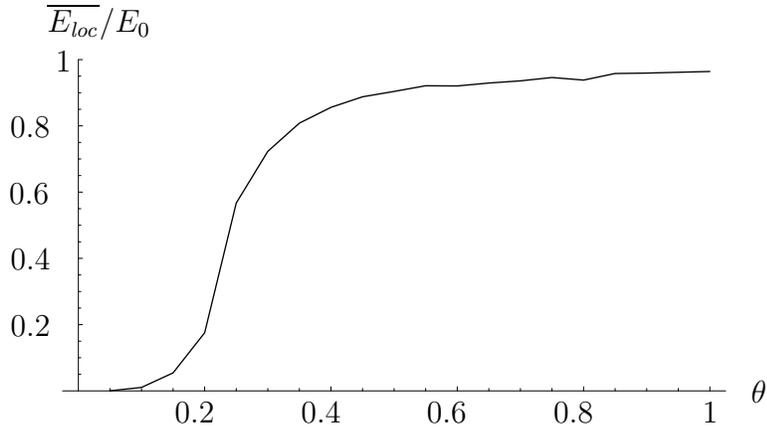,width=0.75\textwidth}
\caption{The energy left in the first four oscillators after the initial
transient time apparent in figs.\ \ref{decay1}-\ref{decay4}, obtained
by averaging $E_{\it loc}(t)/E_{\it loc}(0)$ from $t=30$ to $t=50$, 
for varying $\theta$.}
\label{transient}
\end{center}
\end{figure}

While the energy seems to always dissipate, there are apparently 
significant differences between large and small $\theta$. For
small $\theta$, the system dissipates most of the energy at
early times. For large $\theta$, the system loses a small fraction
of its initial energy rather rapidly at early times, and after
a certain ``transient'' time the energy dissipation rate approaches
a constant but rather small value. To see this more clearly, we estimate
the energy that is left after the transient time by averaging
$E_{\it loc}(t)$ over a certain late time interval, and plot the result 
as a function of $\theta$. The result is shown in fig.\ \ref{transient}.

We see that the turnover happens roughly at $\theta\approx 0.23$. This
agrees rather well with the estimated critical $\theta_c$ below
which the unstable NC soliton is expected to disappear. From 
\cite{djn}, the most naive estimate is
\begin{equation}
\theta_c \sim \left.\frac{\lambda}{|V'(\lambda)|}
\right|_{V''(\lambda)=0}
\approx 0.26
\eqlabel{naive}
\end{equation}
for the potential \eqref{pot}. We also note for comparison that
the period of linear  oscillations is $T_0= 2\pi/\omega_0\approx 1.1$,
where $\omega_0^2=V''(0)$.

In the remainder of the paper, we will compute the rate of energy 
dissipation to a first approximation in the coupling $1/\theta$.

\section{The Setup}
\label{estimate1}

We start from the classical Hamilton function
\begin{equation}
H = \sum_{n=0}^{\infty} \bigl[ {\frac12} \kappa_n^2
+ V(\lambda_n)
+ \frac n\theta (\lambda_n-\lambda_{n-1})^2
\bigr] \,,
\eqlabel{hamilton2}
\end{equation}
describing an infinite collection of oscillators, labeled by $n=0,1,
\ldots$, with $n$-dependent coupling. We want to understand the late time 
behavior at large $\theta$, if at $t=0$ we excite the zeroth oscillator 
highly into the non-linear regime. The numerics indicate that all other
oscillators remain weakly excited during the decay process, so that we 
may linearize around the minimum of $V$. Furthermore, for large $\theta$,
the zeroth oscillator is weakly coupled to all the other ones (the
other oscillators are strongly coupled as $n\to\infty$). Accordingly, we 
split the Hamilton \eqref{hamilton2} into three pieces
\begin{equation}
H = H_0 + H_{\rm cont} + H_{\rm coup}\,,
\end{equation}
where $H_0$ describes the zeroth oscillator that we want to excite,
\begin{equation}
H_0 = \frac12 \kappa_0^2 + V(\lambda_0) \,,
\end{equation}
$H_{\rm cont}$ describes the remaining oscillators linearized around
the minimum, which for convenience we assume to be at $\lambda_{\it MIN}=0$,
\begin{equation}
H_{\rm cont} = \sum_{n=1}^\infty \bigl[{\frac12} \kappa_n^2
+ \frac12 \omega_0^2 \lambda_n^2
+ \frac{n+1}\theta (\lambda_{n+1}-\lambda_n)^2 \bigr] \,,
\eqlabel{Hcont}
\end{equation}
and $\omega_0^2=V''|_{\it MIN}$. Finally, $H_{\rm coup}$ describes
the coupling between $H_0$ and $H_{\rm cont}$
\begin{equation}
H_{\rm coup} = \frac 1\theta (\lambda_1-\lambda_0)^2 \,.
\end{equation}
At this level, the equations of motion are
\begin{equation}
\begin{split}
\ddot\lambda_0 + V'(\lambda_0) &= \frac 2\theta (\lambda_1-\lambda_0) \\
\ddot\lambda_1 + \omega_0^2 \lambda_1 - \frac2\theta\bigl[2(
\lambda_2-\lambda_1)\bigr] &= -\frac2\theta(\lambda_1-\lambda_0) \\
\ddot\lambda_n + \omega_0^2 \lambda_n - \frac2\theta\bigl[(n+1)(
\lambda_{n+1}-\lambda_n) - n(\lambda_n-\lambda_{n-1})\bigr] &= 0 
\qquad\qquad\qquad\text{for $n\ge 2$}
\end{split}
\eqlabel{eqmot2}
\end{equation}

Our strategy is as follows. We intend to treat the RHS of the first two
equations in \eqref{eqmot2} as a perturbation. Without this term, the first
line can be trivially integrated. By our assumptions on the potential $V$,
the solution, which we call $\tilde\lambda_0$, will be periodic, and we call 
the corresponding frequency $\alpha=\frac{dH_0(J_0)}{dJ_0}$, where $J_0$
is the action variable. We then feed this oscillation into the last two 
equations of \eqref{eqmot2}, which we write as
\begin{equation}
\ddot\lambda_n + \omega_0^2\lambda_n + \frac2\theta (A\lambda)_n =
\frac2\theta\delta_{n,1} \tilde\lambda_0 \,, 
\eqlabel{radiate}
\end{equation}
where $A$ is the linear operator
\begin{equation}
(A\lambda)_n = -(n+1)(\lambda_{n+1}-\lambda_n) + n(\lambda_n-\lambda_{n-1})\,,
\eqlabel{Aop}
\end{equation}
supplemented with appropriate boundary conditions (see below). We can
view \eqref{radiate} as a radiation problem with $\tilde\lambda_0$ as
the source. It is then natural to decompose $\tilde\lambda_0$ into Fourier
modes,
\begin{equation}
\tilde\lambda_0 (t) = \sum_{m=-\infty}^\infty a_m \ee^{\ii m\alpha t} \,,
\eqlabel{fourier}
\end{equation}
and to first compute the energy radiated into the continuum modes of $A$ by
a harmonic source. Summing over all Fourier modes, this will give an
estimate for the rate of energy loss of the zeroth oscillator due to the
coupling to the continuum modes. The expectation is that for large
$\theta$, the energy loss per period is small, so that the coupling to
the continuum essentially amounts to adding a friction term for $H_0\,$, 
with dissipation rate depending on the remaining energy. One can in 
principle study the resulting one-dimensional dissipative system, but 
we have not done this here. On the other hand, for small $\theta$, in 
particular $\theta<\theta_c$, the energy loss per period will become 
comparable to $H_0\,$. In fact, this will give an estimate for the critical 
$\theta_c$. 

Most of what follows is concerned with the operator $A$. In the following
section \ref{A}, we will diagonalize $A$, which means finding the
generalized eigenfunctions and in particular their asymptotics at $n\to\infty$.
This will allow in section \ref{estimate2} to compute the power radiated
into the continuum by a harmonic source. Finally, we estimate the Fourier 
coefficients of $\tilde\lambda_0$ to obtain a decay rate of the soliton.

\section{The operator $A$}
\label{A}

The operator $A$ defined by \eqref{Aop} has the form of a discrete version 
of the familiar class of Sturm-Liouville operators, and we will draw on some 
of the theory pertaining to the latter (see, for instance, \cite{richtmyer}).
First, however, we will obtain some elementary results on the nature of
the spectrum of $A$.

\subsection{The spectrum of $A$ is unbounded and absolutely continuous}

We have
\begin{proposition}
The spectrum of $A$ is equal to $[0,\infty)$ and purely absolutely
continuous.
\end{proposition}
To prove the first part of the statement, we recall that the spectrum
of $A_N$,
\begin{equation}
(A_N\lambda)_n = N(2\lambda_n-\lambda_{n+1}-\lambda_{n-1})
\end{equation}
on $l_2(\zet)$ is equal to $[0,4N]$. The spectrum is absolutely 
continuous and the generalized eigenvectors are $\lambda_n = \ee^{\ii k n}$ 
with spectral parameter $\kappa = 2N(1-\cos k)$.

Now, for fixed $\kappa\in[0,\infty)$, we consider $N\in\zet_{\ge0}$ large 
enough such that $\kappa=2N(1-\cos k)$ has a real solution $k(N)>0$. For 
given $N,M\in\zet$, we can find a normalized 'almost-eigenvector' $\lambda=
(\lambda_n)\in l_2(\zet_{\ge 0})$ with support in $[N,N+M]$ and 
\begin{equation}
||(A_N-\kappa)\lambda||\le \const N\Bigl(\frac{k(N)}{M} + \frac {1}{M^2}\Bigr)\,,
\end{equation}
where the constant is independent of $N$ and $M$. (For instance, we can take 
$\lambda_n= f(n)\ee^{\ii k(N) n}$, where $f$ is normalized, has support in 
$[N,N+M]$, and is sufficiently smooth.) We also have an estimate on the 
first derivative $\lambda'$, $(\lambda')_n=\lambda_n-\lambda_{n+1}$, \eg,
\begin{equation}
||\lambda'|| \le \const \Bigl( k(N) + \frac 1M\Bigr) \,.
\end{equation}
Then, using
\begin{equation}
\bigl((A-A_N)\lambda\bigr)_n=\frac{n-N}{N}(A_N\lambda)_n+
\lambda_n-\lambda_{n+1}\,,
\end{equation}
we find
\begin{equation}
||(A-\kappa)\lambda||\le 
\const \Bigl[
\frac{M \kappa}{N} + k(N) + \frac 1M + \Bigl(\frac MN+1\Bigr)
\Bigl( \frac {N k(N)}{M} + \frac{N}{M^2}\Bigr)\Bigr]\,.
\end{equation}
For large $N$, we have $k(N)\sim (\kappa/N)^{1/2}$, and we let 
$M\sim N^{3/4}$. Thus,
\begin{equation}
||(A-\kappa)\lambda||\le
\const \bigl[\kappa N^{-1/4}+ \kappa^{1/2} N^{-1/2} + N^{-3/4}
+\kappa^{1/2} N^{-1/4} + N^{-1/2} \bigr]\,.
\end{equation}
Making $N$ large, the Weyl criterion implies that $\kappa$ is in the spectrum 
of $A$. Positivity of $A$ is trivial.

To prove absolute continuity, we use the so-called positive commutator 
method or Mourre theory \cite{mourre}. The point of Mourre theory is that one 
can prove absolute continuity of parts of the spectrum of a self-adjoint operator 
such as $A$ by simply exhibiting another self-adjoint operator $K$ such that the 
commutator $\ii[A,K]$ is positive on the part of the spectrum in question.

\begin{proof}[A positive commutator]
We write
\begin{equation}
A = 2n+1 - nT_+ - T_-n \,,
\end{equation}
where $(n\lambda)_n = n\lambda_n$, $(T_+\lambda)_n=\lambda_{n-1}$, 
and $(T_-\lambda)_n=\lambda_{n+1}$. We have
\begin{equation}
[n,T_\pm] = \pm T_\pm \,, \qquad [T_+,T_-] = -\delta_{n,0} \,.
\end{equation}
Define $K=\frac\ii2(nT_+ - T_-n)$ and find
\begin{equation}
\begin{split}
\ii[A,K] &= -[2n+1 - nT_+ - T_-n, {\textstyle\frac12}(nT_+-T_-n)] \\
&= - n[n,T_+] + [n,T_-]n  \\ &\qquad\qquad-{\textstyle\frac12}\bigl(
nT_-[T_+,n] + [n,T_-]nT_+ - [T_-,n]T_+n - T_-n[n,T_+]\bigr) \\
&= - nT_+ - T_-n - {\textstyle\frac12}\bigl( 
-nT_-T_+ - T_-nT_+ - T_-T_+n -T_-nT_+\bigr) \\
&= 2n+1 - nT_+ - T_-n = A \ge 0 \,.
\end{split} 
\end{equation}
This shows absolute continuity away from $0$. But $0$ is not an eigenvalue, 
because any solution of $n(\lambda_n-\lambda_{n+1}) - (n+1)(\lambda_{n+1}-
\lambda_n)=0$ is constant and either trivial or nonnormalizable.
\end{proof}

We now turn to a more formal analysis of the discrete Sturm-Liouville
operator $A$.

\subsection{Boundary Conditions}

$A$ is naively a second order difference operator, and it might seem that we
need to specify boundary conditions at $n=0$. However, from the definition
\eqref{Aop}, it is obvious that $n=0$ is a {\em singular point} of the difference
operator (the order of the operator jumps there). Therefore, {\em we do not 
need to specify boundary conditions}, if we consider the operator on 
$l_2(\{0,1,\ldots\})$. We will denote this operator (called $A$ in the
previous subsection) acting on $l_2(\zet_{\ge 0})$ as in \eqref{Aop} {\em 
without boundary conditions}, by $A_0$. An alternative definition of $A_0$ 
would be by a quadratic form.

For our radiation problem \eqref{eqmot2}, \eqref{radiate}, we want to treat
$\tilde\lambda_0$ as an inhomogeneous term added to a homogeneous linear
equation (the LHS of \eqref{radiate}). In order for \eqref{radiate} to 
coincide with \eqref{eqmot2}, we have to define $A$ acting in $l_2(\{1,2,
\ldots\})$ as in \eqref{Aop}, with {\em in addition the boundary 
condition}
\begin{equation}
\lambda_1-\lambda_0 = 0 \,.
\eqlabel{bouco}
\end{equation}

This boundary condition makes $A$ self-adjoint. But note that, as we review
below, the solution of the radiation problem (computation of the resolvent
of $A$) requires the knowledge of the solutions of the finite difference 
equation with arbitrary boundary conditions. Also note that the particular 
form of the boundary condition \eqref{bouco} is a result of the way we 
have separated the perturbation in \eqref{eqmot2}. We could, for instance, 
also have left $\frac 2\theta\lambda_1$ on the LHS of the equations. Our
results should not depend on this choice.

\subsection{The generalized eigenfunctions}
\label{eigenfunctions}

We know already that the spectrum of $A_0$ is $[0,\infty)$ and absolutely 
continuous. The same statement holds obviously for $A$. Here, we want to
complement this result by explicitly diagonalizing $A$. From the spectral
theory of Sturm-Liouville operators, it is well-known that we will need to
know the two linearly independent solutions of
\begin{equation}
(A\lambda)_n = \kappa \lambda_n
\eqlabel{eigen}
\end{equation}
for arbitrary $\kappa\in\complex$. These two solutions will not necessarily
be $l_2$ normalizable and satisfy linearly independent boundary conditions 
at $n=1$. Specifically, we call the two solutions of \eqref{eigen}, 
$\lambda_n=\Phi_\kappa(n)$ and $\lambda_n=\Psi_\kappa(n)$, satisfying the 
boundary conditions
\begin{align}
\Phi_\kappa(0) &= 1 & \Phi_\kappa(1) &= 1-\kappa
\eqlabel{phibouco}\\
\Psi_\kappa(0) &=0 & \Psi_\kappa(1) &= 1 \,.
\eqlabel{psibouco}
\end{align}
The reason for choosing \eqref{phibouco} was that $\Phi_\kappa$ then also
is a solution of $A_0\Phi_\kappa=\kappa\Phi_\kappa$, but this turns out
to be unimportant. The reason for choosing \eqref{psibouco} as boundary 
conditions for $\Psi_\kappa$ is that in this way, the {\it Wronskian} of
the finite difference equation, which is given in the standard Sturm-Liouville
fashion, is equal to one,
\begin{equation}
\begin{split}
W(\Phi_\kappa,\Psi_\kappa) &= (n+1)\bigl(\Phi_\kappa(n) \Psi_\kappa(n+1) - 
\Psi_\kappa(n)\Phi_\kappa(n+1)\bigr) \\
&= {\rm const.} = 1 \,.
\end{split}
\eqlabel{wronskian}
\end{equation}

We note that the equation \eqref{eigen} has been studied in \cite{acatrinei}. 
It seems, however, that one of the functions given in \cite{acatrinei} is 
incorrect, and moreover, the asymptotics of the solutions in the regime where we will
need them have not been worked out. Quite a bit of work on \eqref{eigen} shows 
that the two solutions are as follows.
\begin{align}
\Phi_\kappa(n) &= \sum_{k=0}^n \frac{(-\kappa)^k}{k!}\binom{n}{k} 
\eqlabel{Phi} \\
\Psi_\kappa(n) &= \sum_{k=0}^n \frac{(-\kappa)^k}{k!}\biggl[
\sum_{s=0}^k \binom ks\binom ns (H_{n-s} + H_{k-s} -2 H_s) \biggr]\,,
\eqlabel{Psi}
\end{align}
where $H_k$ are the ``harmonic numbers''\footnote{$H_0=0$}
\begin{equation}
H_k = \sum_{m=1}^k \frac 1m \,.
\end{equation}
It is trivial that \eqref{Phi} and \eqref{Psi} are entire functions
of $\kappa\in\complex$.

We see from \eqref{Phi}, \eqref{Psi} that $\Phi_\kappa(n)$ is a polynomial 
in $\kappa$ of order $n$, while $\Psi_\kappa(n)$ is a polynomial of order 
$n-1$ (it is easy to see that the coefficient of $(-\kappa)^n$ vanishes). 
In fact, $\Phi_\kappa(n)$ is nothing but a {\it Laguerre polynomial}, which 
is also a particular {\it confluent hypergeometric function}. Namely,
\begin{equation}
\Phi_\kappa(n) = L_n(\kappa) = {}_1F_1(-n,1;\kappa) \,.
\eqlabel{conf}
\end{equation}
A similar, but more complicated statement can be made for $\Psi_\kappa(n)$,
see below.

\subsection{Asymptotics}

We wish to determine the asymptotic behavior of $\Phi_\kappa(n)$ and 
$\Psi_\kappa(n)$ for $\kappa$ fixed, and $n\gg|\kappa|$ large, for 
$\kappa$ in the positive half plane $\complex_+:=
\big\{\kappa\in \complex\big|\Re(\kappa)>0 \big\}$. It was noticed before 
that for large $n$, the operator $A$, eq.\ \eqref{Aop}, can be approximated 
by a differential operator, $A\approx -\frac 1{2r}\del_r r\del_r$, where 
$n=r^2/2$. Let us also recall that the {\it Bessel functions} of order zero,
\begin{align}
J_0(x) &= \sum_{k=0}^\infty \frac{(-x^2/4)^k}{(k!)^2} 
\eqlabel{J0} \\
Y_0(x) &= \frac 2\pi \Bigl[ J_0(x) (\ln\frac x2+\gamma) - 
\sum_{k=1}^\infty \frac{(-x^2/4)^k}{(k!)^2} H_k\Bigr] \,,
\eqlabel{Y0}
\end{align}
where $\gamma=\lim_{n\to\infty} (H_n - \ln n)$ is the {\it Euler-Mascheroni}
constant, are two linearly independent solutions of {\it Bessel's 
differential equation}
\begin{equation}
g'' + \frac 1x g' + g = 0 \,.
\eqlabel{bessel}
\end{equation}
We choose the standard branch for the logarithm in \eqref{Y0}. Furthermore, 
it is easy to see that both \eqref{J0} and the series appearing in 
\eqref{Y0} are entire functions of $x\in\complex$. Hence, it is clear that 
$J_0(x)$ and $Y_0(x)$ are analytic in $\complex_+$.

Comparing \eqref{eigen} with \eqref{bessel}, we expect an asymptotic
expansion of $\Phi_\kappa(n)$ and $\Psi_\kappa(n)$ in terms of the Bessel
functions $J_0(x)$ and $Y_0(x)$, with $x= \sqrt{4 n\kappa}$. Here, and in 
all similar cases considered below, we choose the branch of the square root 
for which the signs of $\Im(x)$ and $\Im(\kappa)$ are equal.

\subsubsection{Asymptotics of $\Phi_\kappa(n)$}

For $\Phi_\kappa$, such an asymptotic expansion is indeed known (see, \eg,
\cite{erdelyi}). It can be derived as follows. Because of \eqref{conf}, 
$\Phi_\kappa(n)=L_n(\kappa)$ satisfies the {\it Laguerre differential 
equation}
\begin{equation}
\kappa f'' + (1-\kappa)f' + n f = 0 \,,
\eqlabel{laguerre}
\end{equation}
where differentiation is with respect to $\kappa$. If we make the usual 
substitution $f(\kappa)=\ee^{\kappa/2} g(\kappa)$ with $\kappa= x^2/4n$ 
in eq.\ \eqref{laguerre}, we obtain
\begin{equation}
g'' + \frac 1x g' +\Big(1- \frac{x^2}{(4n)^2}+\frac{1}{2n}\Big) g = 0 \,,
\eqlabel{lagbes}
\end{equation}
where differentiation is with respect to $x$.

For $|x|\ll n$, which is equivalent to $n\gg|\kappa|$, this reduces to 
\eqref{bessel}, up to a perturbation of order $\calo(\frac{|\kappa|}{n})$. 
Thus, in this limit, any solution $f(\kappa)$ of \eqref{laguerre} is 
asymptotic to a linear combination of Bessel functions $J_0(x)$ and $Y_0(x)$, 
where the deviation from the exact solution is an asymptotic series in 
powers of $\frac{|\kappa|}{n}$. The particular linear combination 
at hand can be determined from the behavior at $x,\kappa\to 0$.
 
{} From the expressions for $\Phi_\kappa(n)$ in eq.\ \eqref{Phi} and $J_0(x)$ 
in eq.\ \eqref{J0} it is easy to see that 
\begin{equation}
\lim_{n\rightarrow\infty}\Phi_{\kappa_n}(n)=J_0(x) \,,
\end{equation}
where $\kappa_n:=\frac{x^2}{4n}$, for any fixed value of $x\in \complex_+$.
Thus, by what has been stated above, it follows that for any fixed   
$\kappa\in \complex_+$
and $n\gg|\kappa|$ large,
\begin{equation}
\Phi_\kappa(n) \sim \ee^{\kappa/2} J_0(x(n,\kappa)) \,,
\eqlabel{Phiasym}
\end{equation}
with $x(n,\kappa)=\sqrt{4n\kappa}$.
 
\subsubsection{Asymptotics of $\Psi_\kappa(n)$}

The asymptotics of $\Psi_\kappa(n)$ for large $n\gg|\kappa|$ requires more 
work. It is natural to expect a connection to the {\it second solution} of
Laguerre's differential equation \eqref{laguerre}. Surprisingly, it turns 
out that this solution is nowhere to be found in the usual references,
including \cite{erdelyi,grary,abste}. The point is that eq.\ 
\eqref{laguerre} is a very special case of the  confluent hypergeometric
equation in which both parameters are integer, and the ``logarithmic 
solution'' given in the references reduces to the ordinary one in this case.
One has to go back to \cite{maob} to find that a second solution is given
by
\begin{multline}
K_n(\kappa) = L_n(\kappa) \ln\kappa + \sum_{k=1}^n \frac{(-\kappa)^k}{k!}
\binom{n}{k} (H_{n-k} - 2H_k - H_n) + \eqlabel{Kn} \\
+(-1)^n\sum_{k=n+1}^\infty \frac{n!}{k!} \frac{(k-n-1)!}{k!} \kappa^k \,,
\end{multline}
which one can easily verify by direct substitution, and where we choose
the standard branch of the logarithm. It is easy to see that both series 
on the RHS of \eqref{Kn} define entire functions of $\kappa\in\complex$.  

Before we relate $K_n(\kappa)$ to $\Psi_\kappa(n)$, let us determine its
asymptotics for large $n\gg|\kappa|$, for $\kappa\in \complex_+$.
We use the same logic as for $L_n(\kappa)$ and first consider the limit
$n\rightarrow\infty$, $\kappa\to0$, with $x^2= 4\kappa n$ fixed. Defining 
$\kappa_n:=\frac{x^2}{4n}\,$, we easily see that
\begin{equation}
\begin{split}
K_n\big(\kappa_n\big) &\sim 
L_n(\kappa_n)
\ln \kappa_n - 2 \sum_{k=1}^\infty \frac{(-\kappa_n n)^k}{(k!)^2}
H_k   \\
&\sim \pi Y_0(x)-J_0(x)(\ln n +2\gamma)
\qquad\text{for $n\rightarrow\infty$}\,.
\end{split}
\end{equation}
The same arguments that were used for $L_n(\kappa)$ then imply that for 
any $\kappa\in \complex_+$, and $n\gg|\kappa|$ large,
\begin{equation}
\qquad\;
K_n(\kappa) \sim \pi\ee^{\kappa/2}Y_0(x(n,\kappa))
-\ee^{\kappa/2} J_0(x(n,\kappa)) (\ln n+2\gamma)\,,
\eqlabel{Knasym}
\end{equation}
where $x(n,\kappa)=\sqrt{4 \kappa n}$.

To relate $\Psi_\kappa(n)$ and $K_n(\kappa)$, we rewrite the sums in 
\eqref{Psi} in terms of $\tilde s= k-s$ and $\tilde k= k-\tilde s=s$. 
This yields
\begin{equation}
\Psi_\kappa(n) = \sum_{\tilde s=0}^n 
\frac{(-\kappa)^{\tilde s}}{\tilde s!} \biggl[
\sum_{\tilde k=0}^{n-\tilde s} 
\frac{(-\kappa)^{\tilde k}}{\tilde k!} \binom{n}{\tilde k}
(H_{n-\tilde k}+H_{\tilde s}-2 H_{\tilde k}) \biggr]\,.
\end{equation}
Using this, we claim that for $\kappa\in \complex_+$ and $n$ large,
\begin{equation}
\Big|\Psi_\kappa(n) - \sum_{\tilde s=0}^n 
\frac{(-\kappa)^{\tilde s}}{\tilde s!} \biggl[
\sum_{\tilde k=0}^{n} 
\frac{(-\kappa)^{\tilde k}}{\tilde k!} \binom{n}{\tilde k}
(H_{n-\tilde k}+H_{\tilde s}-2 H_{\tilde k}) \biggr]\Big|
\leq C (8|\kappa|)^n  \frac{n\log n }{n!}  \,,
\eqlabel{Psikappanest}
\end{equation}
for a constant $C$ that is uniform in $n$ and $|\kappa|$. Indeed, the LHS 
of \eqref{Psikappanest} is bounded by
\begin{equation}
\begin{split}
\sum_{\tilde s=0}^n 
\frac{|\kappa|^{\tilde s}}{\tilde s!} \biggl[
\sum_{\tilde k=n-\tilde s+1}^{n} 
\frac{ |\kappa|^{\tilde k}}{\tilde k!} \binom{n}{\tilde k}
\Big|(H_{n-\tilde k}+H_{\tilde s}-2 H_{\tilde k})\Big| \biggr]
&\leq 2^{n+2} \log n \sum_{\tilde s=0}^n 
\sum_{\tilde r=n }^{n+\tilde s}
\frac{|\kappa|^{\tilde r}}{\tilde r!}  
\binom{\tilde r}{\tilde s}\\
&< C n (2|\kappa|)^n  \frac{n\log n }{n!} \sum_{\tilde r=n }^{2n}  
2^{\tilde r}\,,
\end{split}
\end{equation}
since $H_l<2\log n$ for all $1\leq l \leq n$, and 
$\binom{n}{\tilde k}\leq 2^n$. This straightforwardly implies
\eqref{Psikappanest}.

By noticing that the last term in \eqref{Kn} is small in the limit
$n\gg|\kappa|$, with absolute value bounded by
\begin{equation}
\sum_{r=1}^{\infty}\frac{n!}{(n+r)!}\frac{(r-1)!}{(n+r)!}|\kappa|^{n+r}
\leq  \frac{|\kappa|^{n}}{n!} e^{|\kappa|} \,,
\eqlabel{Knest}
\end{equation} 
one then infers that for any $\kappa\in \complex_+$, and large 
$n\gg|\kappa|\,$,
\begin{equation}
\begin{split}
\Psi_\kappa(n) &\sim \ee^{-\kappa} \biggl[\sum_{k=0}^n 
\frac{(-\kappa)^k}{k!}\binom{n}{k} (H_{n-k}-2H_k) \biggr] +
\sum_{s=0}^{\infty} \frac{(-\kappa)^s}{s!} H_s
\biggl[\sum_{k=0}^n \frac{(-\kappa)^k}{k!} \binom{n}{k} \biggr]\\
&\sim \ee^{-\kappa}\bigl[K_n(\kappa) + L_n(\kappa)(H_n-\ln\kappa)\bigr]
+L_n(\kappa) \sum_{s=0}^\infty\frac{(-\kappa)^s}{s!}H_s\,,
\end{split}
\end{equation}
with a small error bounded by the sum of \eqref{Psikappanest} and 
\eqref{Knest}. Using \eqref{Knasym}, we finally find
\begin{equation}
\Psi_\kappa(n)\sim \pi\ee^{-\kappa/2} Y_0(x(n,\kappa)) 
+ \ee^{-\kappa/2} F(\kappa) J_0(x(n,\kappa)) \,,
\eqlabel{Psiasym}
\end{equation}
for $n\gg|\kappa|$, where we have defined
\begin{equation}
F(\kappa) = \ee^\kappa \sum_{s=0}^\infty 
\frac{(-\kappa)^s}{s!} H_s - \gamma - \ln\kappa \,.
\eqlabel{Fkappa}
\end{equation}
It is possible to express $F(\kappa)$ in terms of the incomplete Gamma 
function, but we do not need this here. As a check, one may note that 
the Wronskian of the Bessel functions corresponding to \eqref{bessel} 
is $W(J_0(x(n,\kappa)),Y_0(x(n,\kappa)))=2/\pi$, which after the appropriate 
coordinate transformations confirms $W(\Phi_\kappa(n),\Psi_\kappa(n))=1$.

\subsection{Resolvent and density of states}

We can now compute the resolvent of $A$ and the spectral measure (density of
states) by following the usual Sturm-Liouville procedure. We introduce, 
for $\kappa\in\complex_+$, ${\rm Im}(\kappa)\neq 0$, the functions 
$f_\kappa(n)$ and $g_\kappa(n)$, satisfying $Af_\kappa=\kappa f_\kappa$ 
and $Ag_\kappa=\kappa g_\kappa$, and characterized by the property 
that $f_\kappa$ satisfies the boundary condition \eqref{bouco}, 
$f_\kappa(1)=f_\kappa(0)$, while $g_\kappa$ is normalizable at infinity.
We can express $f_\kappa$ and $g_\kappa$ as linear combinations of 
$\Phi_\kappa$ and $\Psi_\kappa$ and we normalize them such that the 
Wronskian is equal to one,
\begin{equation}
W(g_\kappa,f_\kappa) = 1 \,.
\end{equation}
Using $f_\kappa$ and $g_\kappa$, the resolvent of $A$ can be written as
\begin{equation}
R_\kappa(n,m)= (A-\kappa)^{-1}(n,m) =
\begin{cases} 
f_\kappa(n) g_\kappa(m) \qquad\text{for $n\le m$} \\
g_\kappa(n) f_\kappa(m) \qquad\text{for $n\ge m$}
\end{cases}
\end{equation}
Using \eqref{phibouco} and \eqref{psibouco}, it is easy to see that 
$f_\kappa$ is given by
\begin{equation}
f_\kappa = \Phi_\kappa + \kappa\Psi_\kappa \,.
\eqlabel{fkappa}
\end{equation}
To determine $g_\kappa$, we recall that the Bessel functions $J_0$ and 
$Y_0$ behave at infinity as
\begin{align}
J_0(x)&\mathop{\sim}_{x\to\infty} \sqrt{\frac{2}{\pi x}}\cos
\bigl(x-\frac\pi4\bigr) \eqlabel{J0asym}\\
Y_0(x)&\mathop{\sim}_{x\to\infty} \sqrt{\frac{2}{\pi x}}\sin
\bigl(x-\frac\pi4\bigr) \,. \eqlabel{Y0asym}
\end{align}
Using \eqref{Phiasym} and \eqref{Psiasym}, and $x=2\sqrt{n\kappa}$, we 
see that the normalizable solution, behaving as $J_0(x)\pm\ii Y_0(x)$ for
${\rm Im} (\kappa)\gtrless 0$, is given by
\begin{equation}
g_\kappa = C_\pm \bigl[\pi\ee^{-\kappa}\Phi_\kappa
\pm\ii\Psi_\kappa \mp\ii \ee^{-\kappa} F(\kappa)\Phi_\kappa\bigr] \,,
\eqlabel{gkappa}
\end{equation}
where $C_\pm$ is determined by the requirement on the Wronskian,
\begin{equation}
1=W(g_\kappa,f_\kappa) = C_\pm\bigl[
\kappa\pi\ee^{-\kappa} \mp\ii\kappa\ee^{-\kappa}F(\kappa)\mp\ii\bigr] \,.
\end{equation}

\begin{figure}[ht]
\begin{center}
\psfrag{0}{$0$}
\psfrag{2}{$2$}
\psfrag{4}{$4$}
\psfrag{6}{$6$}
\psfrag{8}{$8$}
\psfrag{10}{$10$}
\psfrag{0.2}{$0.2$}
\psfrag{0.4}{$0.4$}
\psfrag{0.6}{$0.6$}
\psfrag{0.8}{$0.8$}
\psfrag{1}{$\!\!\!\!\!\!\!\!\!1.0$}
\psfrag{rho}{$\rho(\kappa)$}
\psfrag{kappa}{$\kappa$}
\epsfig{file=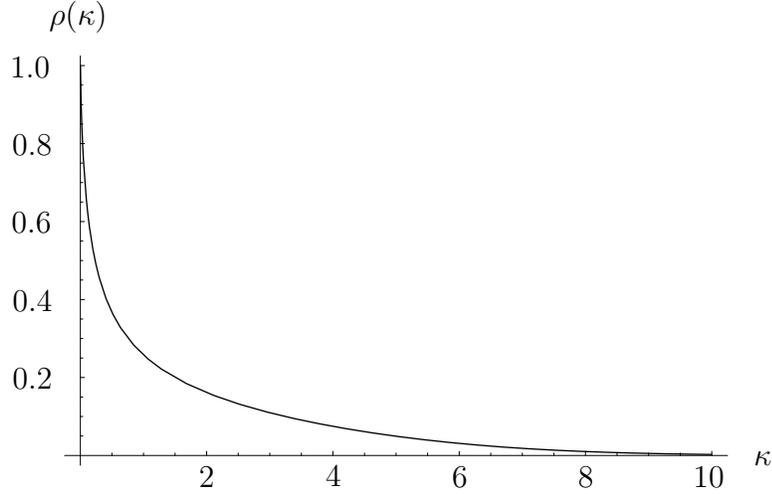,width=0.75\textwidth}
\caption{The density of states \eqref{rho} of the operator $A$ 
\eqref{Aop}.}
\label{rhoplot}
\end{center}
\end{figure}

The density of states or spectral measure $\rho(\kappa)$ of $A$ is
determined from the jump of the imaginary part of the resolvent across 
the real $\kappa$ axis. In the case at hand, one finds in the usual way,
\begin{equation}
\lim_{\epsilon\to 0} \frac  1{2\pi\ii}(g_{\kappa+\ii\epsilon}-
g_{\kappa-\ii\epsilon})= \rho(\kappa) f_\kappa \,,
\end{equation}
which upon using \eqref{fkappa} and \eqref{gkappa} gives, for
$\kappa>0$,
\begin{equation}
\rho(\kappa) = \frac{\ee^{-\kappa}}
{(\kappa\pi\ee^{-\kappa})^2+(1+\kappa\ee^{-\kappa}F(\kappa))^2} \,.
\eqlabel{rho}
\end{equation}
For illustrative purposes, we show a plot of $\rho(\kappa)$ in
fig.\ \ref{rhoplot}.

\section{Estimating the decay rate}
\label{estimate2}

Equipped with the analysis of the linear operator $A$, we now estimate
the decay rate in the approximation explained in section \ref{estimate1}.

\subsection{A harmonic source}
We first study the radiation problem \eqref{radiate} with a harmonic
source.
\begin{equation}
\ddot\lambda_n + \omega_0^2\lambda_n + \frac 2\theta (A\lambda)_n=
\frac2\theta\delta_{n,1}\; a \ee^{\ii\omega t} \,.
\eqlabel{rad}
\end{equation}

There are in principle two ways to treat this problem. One is to 
make an ansatz of the form $\lambda_n(t)= \ee^{\ii\omega t}\mu_n$ 
and to solve for $\mu_n$, imposing outgoing boundary conditions 
at $n\to\infty$. The other is to use the spectral decomposition 
of $\lambda_n(t)$ in eigenmodes of $A$. As a check, we will compute 
with both methods.

\subsubsection{Diagonalization of $A$}

We introduce 
\begin{equation}
\lambda_\kappa(t)=\sum_{n=1}^\infty f_\kappa(n)\lambda_n(t) \,,
\eqlabel{trafo}
\end{equation}
where $f_\kappa$ is given by \eqref{fkappa}. The inverse transformation
involves the density of states,
\begin{equation}
\lambda_n(t) = \int \rho(\kappa)d\kappa\; f_\kappa(n)\;\lambda_\kappa(t)\,.
\eqlabel{inverse}
\end{equation}
Using eq.\ \eqref{trafo} in eq.\ \eqref{rad} yields
\begin{equation}
\ddot\lambda_\kappa +\Bigl(\frac2\theta\kappa+\omega_0^2\Bigr) 
\lambda_\kappa = b\ee^{\ii\omega t} \,,
\end{equation}
where $b=\frac2\theta a f_\kappa(1)$. This is a forced harmonic oscillator 
and can be solved in the usual way. Using \eqref{inverse} to plug the 
solution back into the expression for the energy,
\begin{equation}
E(t) = \sum_{n=1}^\infty \bigl[\frac12\dot\lambda_n^2 +\frac12\omega_0^2
\lambda_n^2\bigr] + \frac1\theta\langle\lambda,A\lambda\rangle \,,
\end{equation}
one finds that at late times, $E$ is proportional to $t$ and the {\it 
radiated power} is given by
\begin{equation}
P=\lim_{t\to\infty} \frac {E(t)}t=\frac\pi{2\theta}\omega a^2\rho(\kappa) 
\eqlabel{Pone}
\end{equation}
where $\kappa=\frac\theta2(\omega^2-\omega_0^2)$ has to be positive,
and $\rho(\kappa)$ is given by \eqref{rho}. The derivation of
\eqref{Pone} is similar to that of Fermi's Golden Rule.

\subsubsection{Modification of boundary conditions}

If we put $\lambda_n(t)=\ee^{\ii\omega t} \mu_n$ into eq.\ \eqref{rad},
we obtain
\begin{equation}
-\omega^2\mu_n + \frac 2\theta (A\mu)_n +\omega_0^2\mu_n =
\frac 2\theta  \delta_{n,1} a \,.
\end{equation}
Thus, for $n\neq 1$, $\mu=(\mu_n)$ must be a linear combination of
$\Phi_\kappa$ and $\Psi_\kappa$, with $\kappa=\frac\theta2
(\omega^2-\omega_0^2)$. We require that $\mu_n$ be {\it outgoing at 
infinity}, which from the behavior of $J_0$ and $Y_0$ means that
$\mu_n$ must behave as $J_0(x)+\ii Y_0(x)$. Hence,
\begin{equation}
\mu_n = D\bigl[ \pi\ee^{-\kappa}\Phi_\kappa(n) + \ii\Psi_\kappa(n)
- \ii\ee^{-\kappa} F(\kappa)\Phi_\kappa(n)\bigr]\,,
\end{equation}
where $D$ is determined by the equation at $n=1$,
\begin{equation}
-\frac2\theta\,2(\mu_2-\mu_1)+(\omega_0^2-\omega^2)\mu_1=
-\frac2\theta\,(\mu_1-\mu_0) = \frac2\theta \,a \,,
\end{equation}
yielding
\begin{equation}
a = D\bigl[\pi\kappa\ee^{-\kappa} -\ii\kappa\ee^{-\kappa}F(\kappa)
-\ii\bigr] \,.
\eqlabel{D}
\end{equation}

To finish the calculation of the energy output, one has to plug the
solution into the expression for the ``{\it energy current}'',%
\footnote{We define $T_{01}$ by the requirement that it satisfy
$\dot T_{00}(n) = - (T_{01}(n)-T_{01}(n-1))$ on a solution of the
equations of motion, where
$T_{00}(n)= \frac12\dot\lambda_n^2+\frac n\theta(\lambda_n-\lambda_{n-1})^2
+\frac12\omega_0^2\lambda_n^2$ is the ``energy density''.}
\begin{equation}
T_{01} = -\frac2\theta(n+1)(\lambda_{n+1}-\lambda_n)\dot\lambda_n \,.
\end{equation}
Using the asymptotics of $\Phi_\kappa$ and $\Psi_\kappa$ and of $J_0$
and $Y_0$, one obtains for the average radiated power
\begin{equation}
P = \overline{T_{01}}(n\to\infty) = \frac{\pi}{2\theta} \omega
\ee^{-\kappa}
|D|^2 \,,
\eqlabel{Ptwo}
\end{equation}
which using \eqref{D}, reproduces exactly \eqref{Pone},
\begin{equation}
P=P(\omega,a,\theta)=\frac\pi{2\theta} \omega a^2 \frac{\ee^{-\kappa}}
{(\kappa\pi\ee^{-\kappa})^2+(1+\kappa\ee^{-\kappa}F(\kappa))^2} \,,
\eqlabel{Pthree}
\end{equation}
with $\kappa=\frac\theta2(\omega^2-\omega_0^2)>0$, and $F(\kappa)$
is given by \eqref{Fkappa}.

\subsection{The decay rate of the NC soliton}

As explained in section \ref{estimate1}, the total decay rate of the NC 
soliton is obtained by decomposing the oscillation $\tilde\lambda_0$ 
into Fourier modes, \eqref{fourier}, using \eqref{Pthree} for each mode, 
and summing over Fourier modes, \ie,
\begin{equation}
P_{\rm tot}(\alpha,\theta) = \sum_{m=-\infty}^\infty P(m\alpha,a_m,\theta)\,,
\eqlabel{total}
\end{equation}
where
\begin{equation}
a_m = \frac{\alpha}{2\pi} \int_0^{2\pi/\alpha}dt\; \ee^{-\ii m\alpha t}
\tilde\lambda_0(t)
\eqlabel{coeff}
\end{equation}
are the Fourier coefficients of $\tilde\lambda_0$. The precise values
of the $a_m$ of course depend on the details of the potential and the 
exact value of the basic frequency $\alpha$ (which, if we enforce energy 
conservation, is not fixed), so that we need an estimate to make further
progress on \eqref{total}.

There are two regimes in which one can estimate the Fourier coefficients.
The first is for almost harmonic oscillations, for which one can use 
perturbation theory to determine the $a_m$. This will be the description 
when, for late times, the zeroth oscillator is in a small vicinity 
of the stationary point, and the soliton has almost completely decayed. 
The second regime is defined by $\alpha\ll\omega_0$. It describes the 
initial stages of the decay of the NC soliton, and it will be a consistent approximation if the 
energy loss per period is much smaller that the total energy in the 
zeroth oscillator, \ie, we require
\begin{equation}
P_{\rm tot}(\alpha,\theta) \ll \alpha H_0\ll\omega_0 H_0 \,.
\eqlabel{condition}
\end{equation}
In this regime, the coefficients \eqref{coeff} can be estimated as 
follows \cite{doug}.

If the energy $H_0$ is exactly equal to the critical value
$V(\lambda_{\it max})$, there is not only the static solution,
but also the separatrix solution that is not periodic, but
rather takes an infinite amount of time to fall off and come
back to the top of the potential. We call this separatrix
$\bar\lambda(t)$. 

For energies close to the critical point of $V$, the oscillation, which 
has a frequency $\alpha\ll\omega_0$, is well approximated by 
the convolution of the separatrix with a sum of delta functions 
separated by $2\pi/\alpha$, which is a much longer time scale than 
the scale typical of $\bar\lambda$, \ie,
\begin{equation}
\tilde\lambda_0(t)\approx \sum_{m=-\infty}^{\infty}
\bar\lambda(t-2\pi m/\alpha) \,.
\end{equation}
The Fourier transform of $\tilde\lambda_0$ is then given by the product 
of the Fourier transform of $\bar\lambda$ with delta functions at 
$m\alpha$, $m\in\zet$. In other words,
\begin{equation}
a_m \approx \frac {\alpha}{2\pi}\int_{-\infty}^{\infty} dt\;
\ee^{-\ii m\alpha t} \bar\lambda(t) = 
\alpha\hat{\bar\lambda}(m \alpha)
\eqlabel{fourierestimate}
\end{equation}
where $\hat{\bar\lambda}$ is the Fourier transform of the separatrix.

Plugging \eqref{fourierestimate} in \eqref{total}, and using that 
$\rho(\kappa)$ is finite everywhere to replace the sum over Fourier
modes by an integral, we obtain the final expression for the decay rate
of the NC soliton,
\begin{equation}
P_{\rm tot}(\alpha,\theta) \approx
\frac{\pi}{2\theta}\alpha \int_{\omega_0}^\infty d\omega\; \omega
|\hat{\bar\lambda}(\omega)|^2 \rho(\kappa) \,,
\eqlabel{final}
\end{equation}
in which $\kappa=\frac 2\theta(\omega^2-\omega_0^2)$.

\begin{figure}[ht]
\begin{center}
\psfrag{0.1}{$0.1$}
\psfrag{0.2}{$0.2$}
\psfrag{0.3}{$0.3$}
\psfrag{0.4}{$0.4$}
\psfrag{0.6}{$0.6$}
\psfrag{0.8}{$0.8$}
\psfrag{1}{$\!\!\!\!1.0$}
\psfrag{theta}{$\theta$}
\psfrag{P}{$P_{\rm tot}(\theta,\alpha)/\alpha V(\lambda_{\it max})$}
\epsfig{file=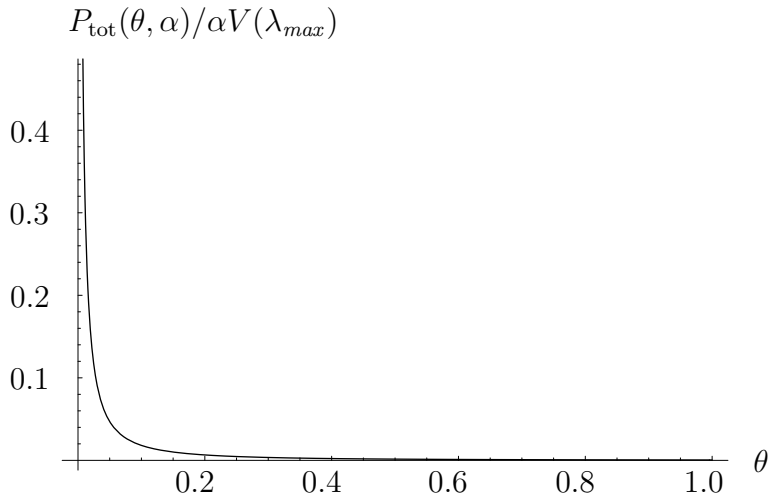,width=0.75\textwidth}
\caption{The energy loss per period of $\tilde\lambda_0$ at
initial stages of the NC soliton decay, as a function of
$\theta$. The potential is of the quartic form \eqref{pot}.}
\label{rate}
\end{center}
\end{figure}

For some special potentials, one can explicitly compute the
separatrix and its Fourier transform. For instance, for
the quartic potential that we used in the numerics,
$V(\Phi)= 4\Phi^2(\Phi-2)^2$, one has%
\footnote{Another simple example that can be solved explicitly
is the cubic potential.}
\begin{align}
\bar\lambda(t) &= \frac{\sqrt{2}}{\cosh 4t} + 1 \,, \\
\intertext{and}
\hat{\bar\lambda}(\alpha) &= \frac{1}{4\sqrt{2}}
\frac{1}{\cosh \frac{\pi\alpha}{8}} \,.
\end{align}
For illustration, we show in fig.\ \ref{rate} the numerical 
evaluation of \eqref{final} for this separatrix. We see that the 
value of $\theta$ at which the energy loss per period becomes 
appreciable is around $\theta\approx 0.2$, in rough agreement 
with the estimate \eqref{naive}.

\section{Conclusions}
\label{conclusions}

In this paper, we have studied some properties of the dynamical system 
\eqref{eqmot}, which is (the radial part of) the classical equations of 
motions for a scalar noncommutative field in $2+1$ dimensions. We have 
in particular analyzed the long time evolution of initial conditions of 
the form (I1) corresponding to unstable noncommutative solitons. We have 
obtained numerical and analytical evidence that the energy is dissipated 
away to infinity in the form of classical radiation. For large, but 
finite values of the noncommutativity parameter $\theta$, 
the decay rate of the soliton 
is exponentially small in $\theta$. Our main technical result is the 
explicit diagonalization of the linear problem. 

Our paper will end here, but we would like to mention a number of open
problems that are worthwhile of further investigation, and to which we
hope to return in the near future.

In the context of NC field theory, there are several other unstable
solitons whose decays one might want to study. We have already mentioned
the scalar soliton decays of type (I2) and (I3). Decays of the type (I2) 
are particularly interesting because they have to proceed through non 
rotationally invariant configurations \cite{djn2}. Solution of this
problem will involve solving the linear problem for nonzero angular
momentum (\ie, finding the ``discrete Bessel functions'' for degree
$l\neq 0$, the function $\Phi_\kappa$ and $\Psi_\kappa$ corresponding
to $l=0$). Furthermore, it is not a priori clear into what decay
products such a soliton will decay. For example, the soliton
corresponding to $\lambda_0^{(0)}=\lambda_2^{(0)}=\cdots=
\lambda_{\it MIN}$, $\lambda_1^{(0)}=\lambda_{\it min}$ could
decay into either the stable soliton corresponding to
$\lambda_0^{(0)}=\lambda_{\it min}$, $\lambda_n^{(0)}=\lambda_{\it 
MIN}$ ($n>0$) or into the global vacuum. An additional question 
is whether there exist rotationally asymmetric stable
solitons, and if so, whether they can be formed as decay products
of the above process. It is natural to ask whether
there is a simple criterion to decide this question.

NC gauge theories, with or without couplings to scalar fields, also 
admit solitons, which in fact have a more direct interpretation in
terms of D-branes in string theory (see \cite{harvey} for a review).
Some of these solitons are unstable and hence can decay. Our solution 
of the linear problem is directly relevant in this context, but the
nonlinearities are rather different, leading to a markedly different
qualitative behavior \cite{wip}.

Also from the point of view of string theory, it would be interesting to
compare our results with the qualitative properties of the decay of the 
lump in $\phi^3$ theory in $1+1$ dimensions, which is another natural
toy model for tachyon condensation. The fluctuations around the lump 
and the expression of the stable vacuum in terms of the fluctuations were 
studied in \cite{zwiebach30}. However, it seems to be unclear whether 
in the time evolution, the field eventually settles down in the stable 
vacuum, and if so, how the energy is dissipated \cite{zwiebach31}. 

Probably the most immediate problem from a mathematical point of view
is to substantiate our qualitative discussion with rigorous estimates
of the decay. As we have mentioned, our system is similar to systems
exhibiting the discrete breather phenomenon \cite{flwi}. One can also
expect a connection to the general theory of resonances in nonlinear
field equations and their radiative decay, studied, for example, in 
\cite{sigal,sowe}. This is the general framework in which we expect 
a more rigorous treatment of our problem.

\begin{acknowledgments}
This work is bearing fruits thanks to the valuable input from
a great number of people, including Tony Zee, Kay Wiese, Tristan 
Maillard, Gian Michele Graf, Stephen Gustafson, Chong-Sun Chu, 
Christopher Herzog, Manfred Salmhofer, Joe Polchinski, Shiraz Minwalla, 
Mark Srednicki, Doug Eardley, and Barton Zwiebach.
J.W.\ would like to thank the CERN Theory Division and the ITP 
at ETH for hospitality during early stages of this work, as 
well as the Perimeter \PI\ Institute during the final stages.
The work of J.W.\ was supported in part by the National Science 
Foundation under Grant No.\ PHY99-07949. T.C.\ was supported by
a Courant Instructorship.

\end{acknowledgments}

\providecommand{\href}[2]{#2}\begingroup\raggedright
\endgroup


\end{document}